\documentclass[prb,showpacs,amsmath,amssymb,superscriptaddress,twocolumn,
floatfix]{revtex4}

\usepackage[english]{babel}
\usepackage{graphicx}
\usepackage{times}
\usepackage{units}

\begin{document}

\title{Fano-Kondo effect in side-coupled double quantum dots at finite
temperatures and the importance of the two-stage Kondo screening}

\author{Rok \v{Z}itko} \affiliation{Jo\v{z}ef Stefan Institute, Jamova 39,
SI-1000 Ljubljana, Slovenia}

\date{\today}


\begin{abstract}
We study the zero-bias conductance through the system of two quantum
dots, one of which is embedded directly between the source and drain
electrodes, while the second dot is side-coupled to the first one
through a tunneling junction. Modeling the system using the
two-impurity Anderson model, we compute the temperature-dependence of
the conductance in various parameter regimes using the numerical
renormalization group. We consider the non-interacting case, where we
study the extent of the departure from the conventional Fano resonance
line shape at finite temperatures, and the case where the embedded
and/or the side-coupled quantum dot is interacting, where we study the
consequences of the coexistence of the Kondo and Fano effects. If the
side-coupled dot is very weakly interacting, the occupancy changes by
two when the on-site energy crosses the Fermi level and a
Fano-resonance-like shape is observed. If the interaction on the
side-coupled dot is sizeable, the occupancy changes only by one and a
very different line-shape results, which is strongly and
characteristically temperature dependent. These results suggest an
intriguing alternative interpretation of the recent experimental
results study of the transport properties of the side-coupled double
quantum dot [Sasaki et al., Phys. Rev. Lett. 103, 266806 (2009)]: the
observed Fano-like conductance anti-resonance may, in fact, result
from the two-stage Kondo effect in the regime where the experimental
temperature is between the higher and the lower Kondo temperature.
\end{abstract}

\maketitle

\newcommand{\expv}[1]{\left\langle #1 \right\rangle}
\newcommand{\korr}[1]{\langle\langle #1 \rangle\rangle}

\renewcommand{\Im}{\mathrm{Im}}
\renewcommand{\Re}{\mathrm{Re}}

\newcommand{\Epsilon}{\mathcal{E}}

\newcommand{\sgn}{\mathrm{sgn}}

\section{Introduction}
\label{sec1}

In condensed-matter physics, the Fano resonance line shape \cite{fano1961,
miroshnichenko2009} is commonly observed in the low-temperature zero-bias
conductance curves of various mesoscopic and nanoscale electronic devices
when the energy of a weakly coupled discrete state is swept across the Fermi
level using gate voltages. A prototype system where Fano physics may be
observed consists of a single quantum dot side-coupled to a quantum
wire \cite{kang2001, bulka2001, aligia2002proetto, torio2002, kobayashi2002,
torio2004filter, aligia2004filter, kobayashi2004, johnson2004, maruyama2004,
sato2005, orellana2006, lobo2006}. The Fano line shape results from an
interference of the quantum amplitudes for the conduction pathway directly
through the quantum wire without passing through the quantum dot and the
indirect conduction pathway via the quantum dot. The first pathway plays the
role of a broad background process, while the second corresponds to a
resonant scattering channel. The conductance as a function of the discrete
state energy level $\epsilon$ is well described by the Fano function
\begin{equation}
\label{eq1}
G(E) = a \frac{\left(E+q\right)^2}{1+E^2} + b,
\end{equation}
where 
\begin{equation}
\label{eq2}
E = \frac{\epsilon - \epsilon_r}{\Gamma}
\end{equation}
is a dimensionless energy that measures the energy difference from a
resonance energy $\epsilon_r$ in units of the resonance half-width $\Gamma$,
$q$ is the Fano parameter given by the ratio of resonant and background
scattering amplitudes, while $a$ and $b$ are some coefficients.
\footnote{In line with the convention followed by the quantum impurity
physics community, the hybridization strength $\Gamma$ is defined in this
work as the resonance half-width. This needs to be contrasted with the
convention where $\Gamma$ is taken to be the full-width (i.e., level
broadening) which is more commonly used by experimentalists. In the letter
case, one has $E=(\epsilon-\epsilon_r)/(\Gamma/2)$.}
This energy dependence holds as long as the background conductance is
constant over the width of the resonance; in experiments, curve fitting with
Fano profile is typically done in an energy window of the order of several
times $\Gamma$.

\begin{figure}[htbp]
\includegraphics[width=5.6cm,clip]{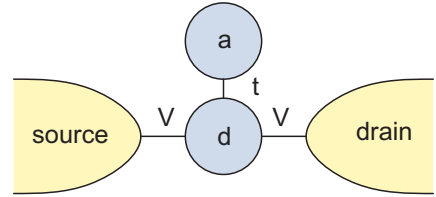}
\caption{(Color online) Schematic representation of the double quantum dot
nanostructure consisting of a quantum dot (d), embedded between the source
and drain electrodes, and a side-coupled quantum dot (a).}
\label{fig0}
\end{figure}

The situation becomes more involved in a related structure consisting of two
quantum dots, the first of which is embedded in the direct conduction
pathway between the source and drain electrodes and the second is
side-coupled to the first one by weak tunnel coupling, as illustrated in
Fig.~\ref{fig0}. This nanostructure, also known as the ``side-coupled double
quantum dot'', allows to study the interference between the direct and
indirect conduction pathways (i.e., the Fano effect), as well as various
correlation effects resulting from the strong electron-electron repulsion
between the confined electrons, in particular the Kondo effect
\cite{glazman1988, cronenwett1998, goldhabergordon1998b,
goldhabergordon1998a, wiel2000, jeong2001, kouwenhoven2001, hewson}. Taking
into account only the electron levels in close vicinity of the Fermi level,
this system can be adequately described at low temperatures using a
two-impurity Anderson model which is a simple extension of the
single-impurity model that is commonly used for studying transport
properties of single ultra-small quantum dots \cite{wiel2003}. The
side-coupled double quantum dots have been intensively studied theoretically
\cite{boese2002, vojta2002, stefanski2004, tanaka2005, cornaglia2005tsk,
sidecoupled, sidecoupled2, karrasch2006, weyrauch2008}, uncovering a
fascinating interplay between many-particle effects and quantum
interference, however the first experimental study of this system has been
performed only very recently (Ref.~\onlinecite{sasaki2009}). Most previous
theoretical works consider the zero-bias conductance in the zero-temperature
limit, while the experiment is decidedly performed at finite temperatures.
It is known that this system is characterized by very low energy scales in
some parameter regimes (in particular in the case of the two-stage Kondo
effect \cite{cornaglia2005tsk, sidecoupled, chung2008}), thus in order to
properly describe its transport properties it is imperative to consider
thermal effects and to calculate the conductance at finite temperatures
using a non-perturbative method. This is the goal of the present work.

This paper is structured as follows. In Sec.~\ref{sec2} the model and the
numerical renormalization group method are presented. The thermal effects
are first studied in the non-interacting model in Sec.~\ref{sec3}, where we
show that at finite temperatures the resonance line shape still has the Fano
form to a good approximation, albeit with temperature-dependent parameters.
In Sec.~\ref{sec4} we then focus on the Fano-Kondo effect in the case where
only the directly embedded quantum dot is interacting and experiences the
Kondo screening; in this case the resonance line shape itself is appreciably
modified by the combined Fano interference and Kondo effect. In
Sec.~\ref{sec5} we finally study the fully interacting case, where
completely different behavior is found. Finally, in Sec.\ref{sec6} the
recent experimental results are examined and compared to theory, leading to
the conclusion that the observed line-shape (an anti-resonance) is
indicative of the occurrence of the two-stage Kondo effect.

\section{Model and method}
\label{sec2}

The Hamiltonian under study takes the following form:
\begin{equation}
\begin{split}
H &= 
\delta_d \left( n_d-1 \right) + 
\delta_a \left( n_a-1 \right) - 
t \sum_\sigma \left( d^\dag_\sigma a_\sigma + \text{H.c.} \right) \\
&+ 
\frac{U_d}{2} \left( n_d-1 \right)^2 + 
\frac{U_a}{2} \left( n_a-1 \right)^2 \\
&+ 
\sum_{k\sigma} \epsilon_k c_{k\sigma}^\dag c_{k\sigma} +
V \sum_{k\sigma} \left( c^\dag_{k\sigma} d_\sigma + \text{H.c.} \right),
\end{split}
\end{equation}
where $n_d = \sum_\sigma d^\dag_\sigma d_\sigma$ and $n_a=\sum_\sigma
a^\dag_\sigma a_\sigma$. The operators $d^\dag_\sigma$ and $a^\dag_\sigma$
are creation operators for an electron with spin $\sigma$ on the embedded
quantum dot $d$ and on the side-coupled dot $a$, see Fig.~\ref{fig0}.  The
on-site electron-electron repulsion is denoted by $U_\alpha$, where $\alpha$
stands for $d$ or $a$. The on-site energies $\epsilon_{\alpha}$ of the dots
are shifted by $U_{\alpha}/2$ in order to define the parameters
\begin{equation}
\delta_{\alpha}=\epsilon_{\alpha}+U_{\alpha}/2,
\end{equation}
which measure the detuning from the particle-hole symmetric case which
corresponds to $\delta_a=\delta_d=0$. The coupling between the dots is
described by the interdot tunnel coupling $t$. The dot $d$ is assumed to
couple to the source and drain electrodes in a symmetric way, thus it
hybridizes with the even-parity combination of electrons from both leads
\cite{glazman1988, pustilnik2004} with the strength 
\begin{equation}
\Gamma_d=\pi \rho V^2,
\end{equation}
which will be assumed to be constant for all energies within the conduction
band of half-width $D$ (equivalently, the density of states in the
conduction band, $\rho$, is assumed to be constant, $\rho=1/2D$).
Throughout this work, $\Gamma_d$ will be fixed to $\Gamma_d/D=0.06$.

For the correct description of correlated regimes in clusters of
quantum dots \cite{izumida2000, borda2003, galpin2005, oguri2005,
mravlje2006, martins2006, tripike, nisikawa2006, zarand2006,
vzporedne, dasilva2006, vzporedne2, karrasch2006, flnfl3, martins2006,
lobos2006, trikotnik, anda2008, dasilva2008ft} it is necessary to use
non-perturbative techniques that take properly into account not only
the local correlation effects on the impurity sites, but also the
charge fluctuations and inter-site spin-spin coupling induced by the
exchange interaction. One such technique is the numerical
renormalization group (NRG) \cite{wilson1975, krishna1975,
krishna1980a, krishna1980b, yoshida1990, oliveira1994, costi1994,
bulla1998, costi2001, hofstetter2000, izumida2005, peters2006,
weichselbaum2007, bulla2008, campo2005, resolution} which involves
three steps: 1) logarithmic discretization of the conduction band, 2)
mapping onto a one-dimensional tight-binding chain with exponentially
decreasing hopping constants, and 3) iterative diagonalization of the
resulting Hamiltonian. The results presented in this work have been
calculated using the NRG discretization parameter $\Lambda=4$ with
$N_z=8$ equidistant values of the twist parameter $z$ using an
improved discretization scheme \cite{resolution, odesolv}, with the
truncation cutoff set at $10 \omega_N$. The expectation values are
calculated in the standard way, while the conductance curves at finite
temperatures are computed using the Meir-Wingreen formula from the
spectral data \cite{costi1994, costi2001, yoshida2009prb}:
\begin{equation}
\label{mwf}
G(T) = G_0 \pi \Gamma_d \int_{-\infty}^{\infty}
d\omega \left( -\frac{\partial f}{\partial \omega} \right)
A_d(\omega, T),
\end{equation}
where $G_0=2e^2/h$ is the conductance quantum (i.e., the conductance
corresponding to full transmission), $f(\omega)=[1+\exp(\beta\omega)]^{-1}$
is the Fermi function where $\beta=1/k_B T$ and the chemical potential has
been fixed at zero energy, and $A_d(\omega,T)$ is the spectral function
on the impurity site $d$.

It should be noted that there is a significant conceptual difference as
regards the usage of the Fano formula in the context of scattering
experiments in atomic physics and the present context of nanoscopic
transport. In atomic physics, the resonance is observed in the scattering
cross-section as a function of the energy of the incoming particle while the
resonance energy is fixed. In transport experiments, the resonance is
observed in the conductance which relates to the scattering of electrons at
the fixed energy of the Fermi level (this is strictly true only at zero
temperature; at finite temperatures, electrons within an energy window of
order $k_B T$ participate in the transport) and the independent variable is
the energy of the side-coupled dot which is swept using gate voltages. In
spite of the seemingly symmetric roles of parameters $\epsilon$ and
$\epsilon_r$ in the expressions in Eqs.~\eqref{eq1} and \eqref{eq2}, there
is a notable difference as regards the role of the background channel. The
background scattering phase shift namely depends on the variable parameter
in the first case, while it is essentially constant in the second case (at
zero temperature). There may be some small variation of the background
scattering phase shift when the parameters of the side-coupled dot are
changed, since the two impurities are coupled and there will be
hybridization effects induced by the side-coupled dot on the embedded dot. 
Nevertheless, this effect is rather small. We therefore conclude that any
significant deviation from the universal Fano line-shape profile may only
result from electron correlation effects or from thermal effects.

Furthermore, it should be remarked that the approach used in this work does
not take into account that at high temperatures the phase coherence
necessary for the Fano interference may be partially lost, since full
quantum coherence is assumed in solving the model \cite{zacharia2001,
kobayashi2004}. This implies that in experiments further reduction of the
resonance amplitude is expected in addition to that found in our
calculations which only take into account the effects due to the thermal
broadening of the electron distribution function in the source and drain
electrodes.

In the following, we set $k_B=1$ and the conductance is always expressed in
units of $G_0$.

\section{Non-interacting case}
\label{sec3}

The Fano formula has been originally formulated for a quantum system in its
ground state. For a non-interacting system ($U_a=U_d=0$), it is thus
expected to describe the zero-temperature conductance essentially exactly
as long as the background remains constant, which is the case if the
Fano resonance is narrower than the Lorentzian spectral peak on the embedded
quantum dot (of half-width $\Gamma_d$). The characteristic energy scale of
the level broadening on the side-coupled quantum dot is
\begin{equation}
\gamma_a(\epsilon_a) = \pi A^0_d(\epsilon_a) t^2,
\end{equation}
where $A^0_d(\omega)$ is the spectral function of the level $d$ in the
absence of the level $a$, which is given by
\begin{equation}
A^0_d(\omega) = \frac{1}{\pi \Gamma_d}
\frac{1}{(\omega-\epsilon_d)^2/\Gamma_d^2+1}
\end{equation}
in the wide-band limit which applies well here for the chosen
$\Gamma_d=0.06D$. For later reference, we also give the full spectral
function of the level $d$ in the presence of the side-coupled dot:
\begin{equation}
\label{adw}
A_d(\omega) = \frac{1}{\pi \Gamma_d} \frac{\Omega^2}
{\left[ 1-(\omega-\epsilon_d)(\omega-\epsilon_a)/t^2 \right]^2
+\Omega^2},
\end{equation}
with $\Omega=(\omega-\epsilon_a)/(t^2/\Gamma_d)$.

The non-interacting case is also a good test for the NRG procedure, since a
comparison with essentially exact results from simple quadrature of the
spectral function is possible. We find that the results (detailed in the
following subsections) are correct within the line-width or, more precisely,
within less than half percent. A similar degree of precision is then also
expected in the interacting case considered later on, since the NRG method
functions equally well for any value of $U$.

\subsection{Symmetric case, $\epsilon_d=0$}
\label{sec3a}

Fixing $\epsilon_d=0$ and taking the value of $A_d^0(\omega)$ at the Fermi
level, $A^0_d(0)=1/\pi\Gamma_d$, we obtain
\begin{equation}
\Gamma_a \equiv \gamma_a(0) = \frac{t^2}{\Gamma_d}.
\end{equation}
The Fano formula applies for $\Gamma_a \ll \Gamma_d$, therefore the range of
validity is expected to be given by $t \ll \Gamma_d$. In fact, for
$\epsilon_d=0$ it turns out that the zero-temperature linear conductance
$G(0)=G_0 \pi \Gamma_d A_d(0)$ reduces {\it exactly} to the Fano formula for
any value of the $t/\Gamma_d$ ratio:
\begin{equation}
G(0) = \frac{E^2}{1+E^2},
\end{equation}
with $E=\epsilon_a/\Gamma_a$. The asymmetry parameter $q$ is clearly zero,
thus the Fano resonance takes the form of a symmetric anti-resonance. 

\begin{figure}[htbp]
\includegraphics[width=7cm,clip]{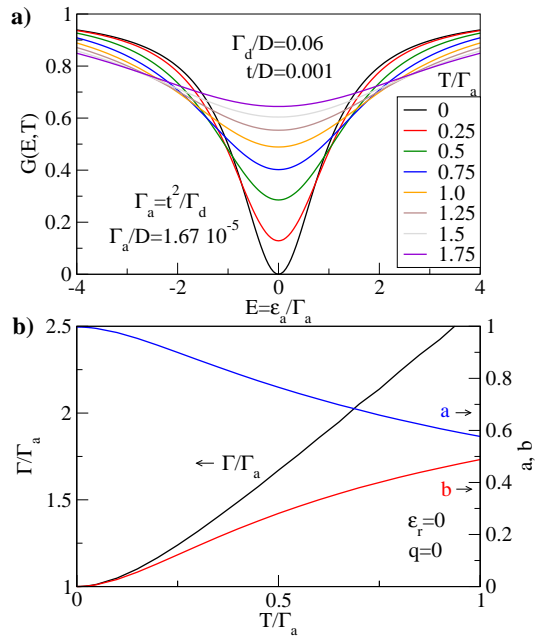}
\caption{(Color online) Non-interacting model with $U_a=U_d=0$, symmetric
case, $\epsilon_d=0$. a) Conductance curves for a range of temperatures.
Fano-resonance-line fits overlap completely with the conductance curves. b)
Fano parameters as a function of the temperature. }
\label{fig1}
\end{figure}

The conductance curves for the non-interacting model with $\epsilon_d=0$ are
shown in Fig.~\ref{fig1}a. At all temperatures, the resonance line-shape can
be described with excellent accuracy using the Fano formula with temperature
dependent parameters which are shown in Fig.~\ref{fig1}b. 
The characteristic temperature scale for the temperature-dependence of the
conductance curves is, as expected, $\Gamma_a$. An approximate expression
for the temperature dependence of the resonance width in the temperature
interval $[0:\Gamma_a]$ is
\begin{equation}
\Gamma = \Gamma_a \left[1 + c_1 (T/\Gamma_a)^2 \right]^{c_2},
\end{equation}
with $c_1=9.2$ and $c_2=0.46$. This temperature-variation rule holds only
in the $t/\Gamma_d \to 0$ limit: in general, the temperature variation also
depends on the $t/\Gamma_d$ ratio.

We have verified that for $\epsilon_d=0$ the Fano curve accurately describes
the finite-temperature conductance even for $t$ of the order of $\Gamma_d$
(results not shown).

\subsection{Asymmetric case, $\epsilon_d \neq 0$}
\label{sec3b}

For $\epsilon_d \neq 0$, the $d$-level Lorentzian is no longer fixed at the
Fermi level, which has a number of significant consequences. The effective
level broadening on the side-coupled dot is reduced and it is given by
\begin{equation}
\Gamma_a \equiv \gamma_a(0) = \frac{t^2}{\Gamma_d} \frac{1}{(\epsilon_d/\Gamma_d)^2+1}.
\end{equation}
Furthermore, the background phase shift is no longer $\pi/2$, thus the
asymmetry parameter $q$ is different from 0. From Eq.~\eqref{adw} it is
clear that the zero-temperature conductance is always zero at
$\epsilon_a=0$, however the actual resonance energy $\epsilon_r$ is actually
shifted away from zero. In fact, within an error of order $t^2/\Gamma_d^2$,
the following Fano parameters hold at $T=0$ for general $\epsilon_d$:
\begin{equation}
\begin{split}
\epsilon_r &= \epsilon_d, \\
\Gamma &= \Gamma_a, \\
q &= \epsilon_d/\Gamma_d, \\
a &= \frac{1}{1+\left( \epsilon_d / \Gamma_d \right)^2}, \\
b &= 0.
\end{split}
\end{equation}

\begin{figure}[htbp]
\includegraphics[width=7cm,clip]{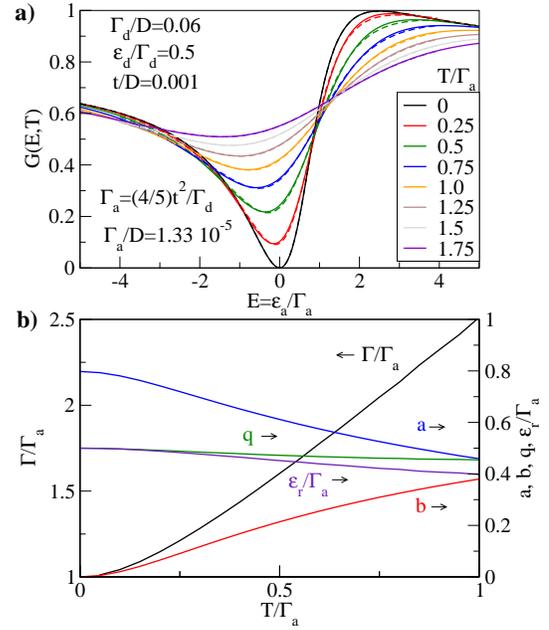}
\caption{(Color online) Non-interacting model with $U_a=U_d=0$,
asymmetric case, $\epsilon_d \neq 0$.
a) Conductance curves for a range of temperatures. Full curves: NRG results.
Dashed lines: Fano resonance line fits.
b) Fano parameters as a function of the temperature.
Parameters are extracted by curve fitting in the energy window which
corresponds to the horizontal axis in the upper subfigure.
}
\label{fig1asym}
\end{figure}

The results of a temperature-dependent calculation for
$\epsilon_d/\Gamma_d=0.5$ are shown in Fig.~\ref{fig1asym}. We observe a
departure from the Fano line-shape which grows as the temperature is
increased; nevertheless, the departure is small and it is still meaningful
to perform curve fitting using the Fano formula. Similar to the symmetric
$\epsilon_d=0$ case, the main effect of finite temperature is to increase
the width of the Fano resonance beyond $\Gamma_a$ and to reduce the
amplitude of the resonance. The asymmetry parameter $q$ is, however, nearly
constant: it reduces from $q=0.5$ at zero temperature to $q=0.47$ at
$T=\Gamma_a$, while in the same temperature interval the amplitude is
reduced by half. The resonance position, as given by $\epsilon_r$, is also
affected rather weakly. Such behavior is characteristic of the $t \ll
\Gamma_d$ limit. For larger $t$, the extracted Fano parameters $q$ and
$\epsilon_r$ are more strongly temperature dependent, however the deviation
from the Fano profile remains small (results not shown).

We conclude, fully in line with the expectations, that in the
non-interacting case the main effect of finite temperature is to reduce the
amplitude and increase the width of the resonance, while the Fano line-shape
is largely preserved.  As we show in the following, this is no longer the
case in the presence of interactions, except in some special limits.

\section{Interacting case with $U_d\neq0$ and $U_a=0$}
\label{sec4}

The occurrence of the Fano resonance in transport problems is always
associated with changes of the charge state of the weakly coupled discrete
state. For this reason, there is a significant difference between the $U_a
\ll \Gamma_a$ and $U_a \gg \Gamma_a$ situations. In the first limit, the
occupancy changes by 2 (by one electron for each spin orientation) as
$\epsilon_a$ crosses the Fermi level, thus the phase change in each spin
channel is $\pi$: this corresponds to the usual Fano-resonance scenario. In
the second limit, the total occupancy changes only by one due to the
electron-electron repulsion preventing the second electron from entering the
quantum dot. This corresponds to a $\pi/2$ phase shift in each spin channel
and, roughly speaking, only one half of the Fano-resonance-like feature is
expected to be seen. In this section we discuss the first case, i.e., the
$U_a \ll \Gamma_a$ limit, which we study by setting $U_a$ to zero.

\begin{figure}[htbp]
\includegraphics[width=7cm,clip]{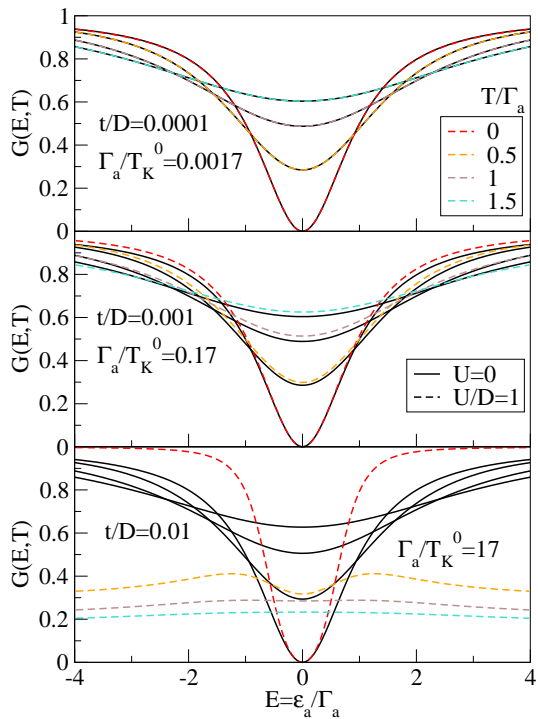}
\caption{(Color online) Resonance curves for non-interacting (full lines)
and interacting models (dashed lines) for a range of scaled temperatures
$T/\Gamma_a$ and for different $\Gamma_a/T_K^0$ ratios. }
\label{fig2}
\end{figure}

When the embedded dot is tuned to its particle-hole symmetric point
($\delta_d=0$), the impurity spectral function $A_d^0$ is pinned at the
Fermi level to the value of $1/\pi\Gamma_d$ irrespective of the value of the
electron-electron repulsion $U_d$.  This is true even for moderate
departures from the particle-hole symmetric point, $|\delta_d| \lesssim
U_d/2$, since the Kondo resonance remains near the Fermi level, unlike the
Lorentzian peak in the non-interacting case. For small $t$, the
characteristic energy scale of the side-coupled quantum dot is therefore
still $\Gamma_a=t^2/\Gamma_d$ as in the symmetric non-interacting case. For
larger $t$ the broadening will, however, eventually become comparable to the
width of the Kondo resonance (of order $T_K^0$). The behavior therefore
depends on the ratio between the hybridization of the side-coupled dot
$\Gamma_a$ and the Kondo temperature of the decoupled embedded dot, $T_K^0$.
If $\Gamma_a \ll T_K^0$, the physics of the embedded dot will not be
affected much by the side-coupled dot (and vice versa), thus the conductance
as a function of $\epsilon_a$ will differ only slightly from the
non-interacting case. This is illustrated in Fig.~\ref{fig2}, upper panel,
where we compare conductance curves for $U_d=0$ and $U_d/D=1$ at
$t/D=0.0001$, where $\Gamma_a/T_K^0=1.7 \times 10^{-3}$ since the Kondo
temperature (Wilson's definition) is $T_K^0=9.7 \times 10^{-5}D$. In this
regime the Kondo effect on the embedded dot in no way affects the Fano
interference process in the relevant range of energies and temperatures
(i.e. several times $\Gamma_a$). Only for very large $\epsilon_a$ and very
high temperature (of the order of $T_K^0$) will the differences become
apparent, however this is outside the parameter regime of interest here.

When $\Gamma_a$ is equal to a considerable fraction of $T_K^0$, we start to
see small quantitative departure from the Fano line shape already for
$\epsilon_a$ and $T$ of the order of $\Gamma_a$, see Fig.~\ref{fig2}, middle
panel. 
This is clearly a consequence of the competition between correlation and
interference effects, combined with further thermal effects. As expected,
the discrepancy grows with increasing $\epsilon_a$ and $T$.

Finally, for $\Gamma_a \gg T_K^0$, the embedded dot is strongly perturbed
by the coupling to the side-dot and the differences become drastic:
at finite $T$, the line-shape differs qualitatively from the Fano form and,
in particular, we observe the emergence of non-uniform $E$-dependence with
broad humps around $E \sim \Gamma_a$.

\begin{figure}[htbp]
\includegraphics[width=7cm,clip]{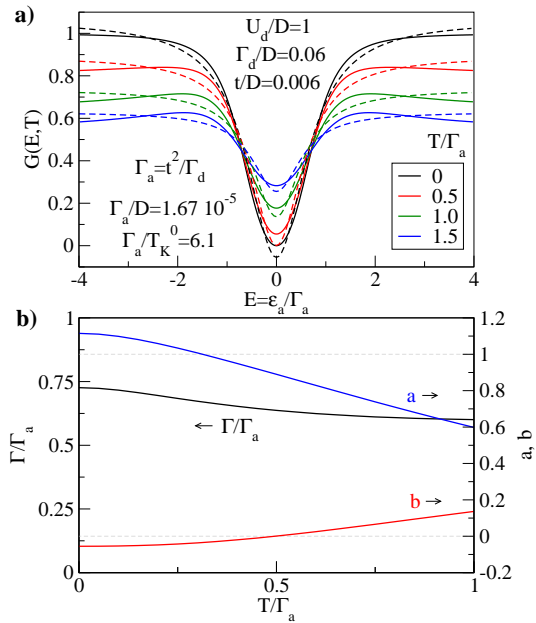}
\caption{(Color online) Interacting model with $U_d \neq 0$ and $U_a=0$.
Symmetric case, $\epsilon_d=0$.
a) Conductance curves for a range of temperatures.
b) Fano parameters as a function of the temperature.
Curve fitting is performed in the energy window which
corresponds to the horizontal axis in the upper subfigure.
}
\label{fig4}
\end{figure}

It is instructive to perform curve fitting on the results where the Fano
resonance is already strongly perturbed,
as shown in Fig.~\ref{fig4}. The fits (shown using dashed lines), performed
in the energy interval $[-4\Gamma_a:+4\Gamma_a]$, are clearly inadequate.
Extracted Fano parameters indicate that the resonance width is significantly
reduced below $\Gamma_a$ and its width even decreases with increasing
temperature, which reflects the situation where the spectral function of the
embedded dot has a significant variation on the energy scale of $\Gamma_a$,
which affects the resonance line-shape for large $E$. It should also be
noted that parameters $a$ and $b$ have ``unphysical'' values $a>1$ and $b<0$
for small $T$. If the curve fitting is performed in a narrower energy
interval $[-\Gamma_a:+\Gamma_a]$, the asymptotic small-$E$ form of the
resonance can be well captured, however the extracted Fano parameters are
factitious and therefore of little use. This demonstrates that in
the presence of strong competition between the Fano interference and Kondo
effect, the curve fitting using the Fano line-shape is not recommended since
the results depend very strongly on the energy window where fitting is
performed and it is thus not advisable to make any inference based on them.

\begin{figure}[htbp]
\includegraphics[width=7cm,clip]{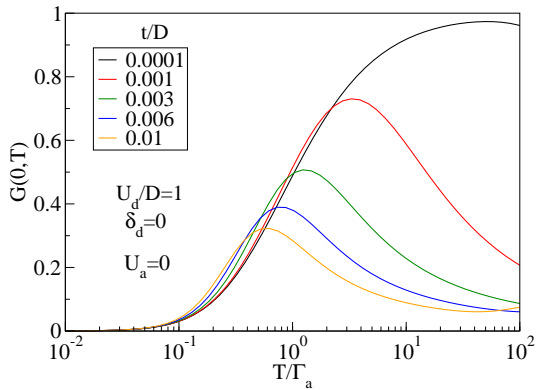}
\caption{(Color online) Temperature-dependence of the conductance for
$\epsilon_a=0$, i.e. at the bottom of the conductance anti-resonance, for a
range of the inter-dot coupling strengths.}
\label{fig3}
\end{figure}

Since there are two different energy scales in the problem ($T_K^0$ and
$\Gamma_a$), the temperature-dependence of the conductance is expected to be
non-monotonic, as shown in Fig.~\ref{fig3} where we plot the conductance at
the bottom of the Fano anti-resonance (i.e. for $\epsilon_a=0$). Such
dependence is a consequence of the competition between the Kondo and Fano
effects. The Kondo effect tends to increase the conductance through the
formation of many-particle resonance at the Fermi level which opens a new
conduction channel through the system. On the other hand, the Fano effect
suppresses the conductance through quantum interference. For $T_K^0 \gg
\Gamma_a$, the conductance first increases at the higher temperature scale
of $T_K^0$ (the temperature dependence being given by the universal Kondo
conductance curve \cite{yoshida2009prb}), then it decreases at
the lower temperature scale set by $\Gamma_a$: see results for $t/D=0.0001$
and $t/D=0.001$ in Fig.~\ref{fig3}. Note that the unitary limit of full
conductance quantum is not achieved at intermediate temperatures even in the
case of well separated energy scales (three orders of magnitude for
$t/D=0.0001$). For larger $t$, the conductance peaks near $T \sim \Gamma_a$,
where both competing effects are equally suppressed by thermal fluctuations.

\begin{figure}[htbp]
\includegraphics[width=7cm,clip]{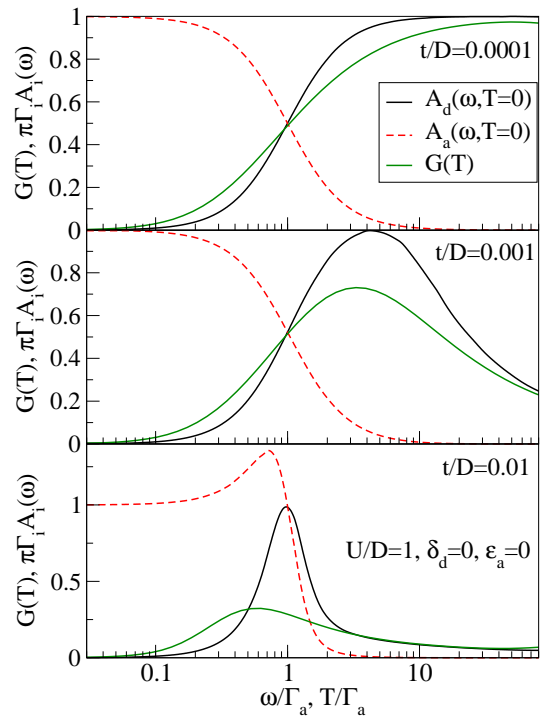}
\caption{(Color online) Comparison of the temperature-dependence of
the conductance $G(T)$ with the energy-dependence of the
zero-temperature spectral function on the embedded dot
$A_d(\omega,T=0)$. For completeness, the spectral function on the
side-coupled dot $A_a(\omega,T=0)$ is also shown. Spectral functions
are rescaled in units of the respective effective hybridisation
strength, i.e., by $1/\pi\Gamma_i$, where $i\in \{a,d\}$. Conductance
curves are rescaled in units of the conductance quantum $G_0=2e^2/h$.
The horizontal axis is rescaled in units of $\Gamma_a=t^2/\Gamma_d$
(note that $\Gamma_a$ is different in each subfigure).}
\label{spec1}
\end{figure}

It is also worthwhile to compare the temperature-dependence of the
conductance $G(T)$ with the energy-dependence of the zero-temperature
spectral function $A_d(\omega,T=0)$, see Fig.~\ref{spec1}. A common
approximation in discussing the transport properties of quantum dot
systems is to assume that the temperature variation of the conductance
simply follows the energy variation of the zero-temperature spectral
function, i.e., $G(T) \approx \pi \Gamma_d A_d(T,0)$. While this is
true on the qualitative level, the numerical results shown significant
qualitative differences: the conductance curves are generally
significantly broader and, accordingly, their maxima are lower. In
particular, while the spectral function often attains values very
close to $1/\pi\Gamma_d$ in some frequency range, the conductance is
not unitary in the corresponding temperature range.

The parameter regime studied in this section corresponds to the model used
in Ref.~\onlinecite{sasaki2009} to obtain the conductance plots featuring an
anti-resonance, which were proposed to explain the experimentally observed
Fano-resonance-like line shapes with small $q$. It can be noticed, however,
that this parameter regime is not appropriate, since it corresponds to a
change of occupancy of the side dot by 2, while from the experimentally
observed honeycomb stability diagram one can deduce that the electron number
changes only by one. We return to this issue in Sec.~\ref{sec6}.

\section{Fully interacting case}
\label{sec5}

\begin{figure}[htbp]
\includegraphics[width=8.5cm,clip]{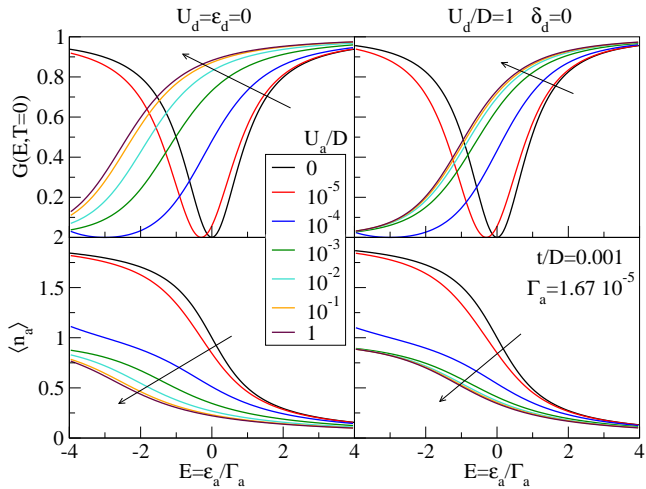}
\caption{(Color online) Conductance curves and level-occupancies at $T=0$
for a range of electron-electron interaction strengths $U_a$. Left panels:
non-interacting embedded impurity. Right panels: interacting embedded
impurity. The arrows show the direction of increasing $U_a$.}
\label{fig6}
\end{figure}

We first study the effect of finite electron-electron repulsion on the
side-coupled dot in the case where the embedded dot itself is
non-interacting, $U_d=0$. In this way we exclude all processes related
to the inter-dot exchange interaction and we may focus on the
path-interference physics. To simplify the discussion, we focus solely
on the symmetric case with $\epsilon_d=0$. As previously discussed, as
long as $U_a \ll \Gamma_a$, the side-coupled dot is essentially
non-interacting and the results are equivalent to those shown in
Sec.~\ref{sec3a}. For larger $U_a > \pi \Gamma_a$, however, the
side-coupled dot will undergo the Kondo effect and its occupancy will
be pinned to 1 for $\epsilon_a$ below several times $\Gamma_a$.
Consequently, the ``resonance'' line-shape in this regime will take
the form of an inverted ``Kondo plateau onset''. The evolution from
the Fano resonance to the Kondo plateau onset is shown in the upper
left panel in Fig.~\ref{fig6}, while the associated level occupancy
curves are shown in the bottom left panel. (See also Fig.~3 in
Ref.~\onlinecite{dasilva2008ft} where a similar setup is considered.)

We now finally discuss the fully interacting case, which is the situation
which is relevant for the actual experimental configuration. The results are
shown in the right panels in Fig.~\ref{fig6}. We find a very similar
evolution in the conductance and occupancy curves with increasing $U_a$ as
in the $U_d=0$ case. The transition from the Fano line shape to the Kondo
plateau-edge shape occurs in this case for lower values $U_a$, however, and
already by $U_a/D=0.01$ we obtain conductance curves which do not change
much as $U_a$ is further increased. This is very important since it implies
that the regime of significantly large $U_a$ occurs for rather small values
of $U_a$, therefore it is not appropriate to model the side-coupled dot as a
non-interacting impurity, since it is unlikely that the electron-electron
repulsion in two similar quantum dots would differ by orders of the
magnitude. We now face a dilemma: assuming a non-interacting side-coupled
dot, we obtain conductance curves which appear to agree with the
experimental results (albeit we know that they correspond to a change of
occupancy by 2, in contradiction with the experimental situation), while
assuming an interacting side-coupled dot, we obtain very different results
which are in strong disagreement with those actually observed.

\begin{figure}[htbp]
\includegraphics[width=7cm,clip]{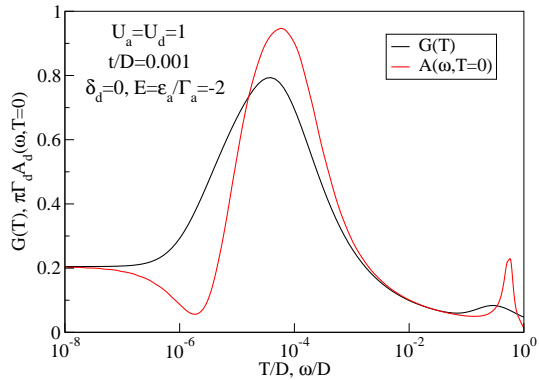}
\caption{(Color online) Comparison of the temperature-dependence of
the conductance $G(T)$ with the energy-dependence of the
zero-temperature spectral function on the embedded dot
$A_d(\omega,T=0)$.}
\label{spec2}
\end{figure}

In Fig.~\ref{spec2} we compare the temperature dependence $G(T)$ with
the energy-dependence of the $T=0$ spectral function
$A_d(\omega,T=0)$. As before (cf.~Fig.~\ref{spec1} and the related
discussion), we observe that in spite of a certain general similarity
between the curves, the differences are notable. In this case, the
differences are even qualitative: while $G(T)$ exhibits two ranges of
logarithmic behavior (the first Kondo screening stage from $10^{-4} <
T <10^{-2}$, and the second Kondo screening stage from $10^{-6} < T <
10^{-4}$), the zero-temperature spectral function is more complex and
even attains a minimum at $\omega \approx 2\times10^{-6}$ before
increasing back to its asymptotic value at the Fermi level. This
complex behavior is {\sl not} reflected in the temperature dependence
of the transport properties. Additional calculations (results not
shown) indicate that $G(T)$ and $A_d(\omega,T=0)$ are very similar
only in vicinity of the particle-hole symmetric point, and even there
only when the two Kondo temperature scales are well separated. This
should serve as a caveat: since $G(T)$ is a convolution of the {\sl
finite-temperature} spectral function, see Eq.~\eqref{mwf}, it is not
expected that, in general, all the details of the energy-dependence of
the zero-temperature spectral function will manifest as similar
variations in the temperature-dependence of the conductance at the
related temperatures.

\section{Relation to the experimental results}
\label{sec6}

\begin{figure}[htbp]
\includegraphics[width=8cm,clip]{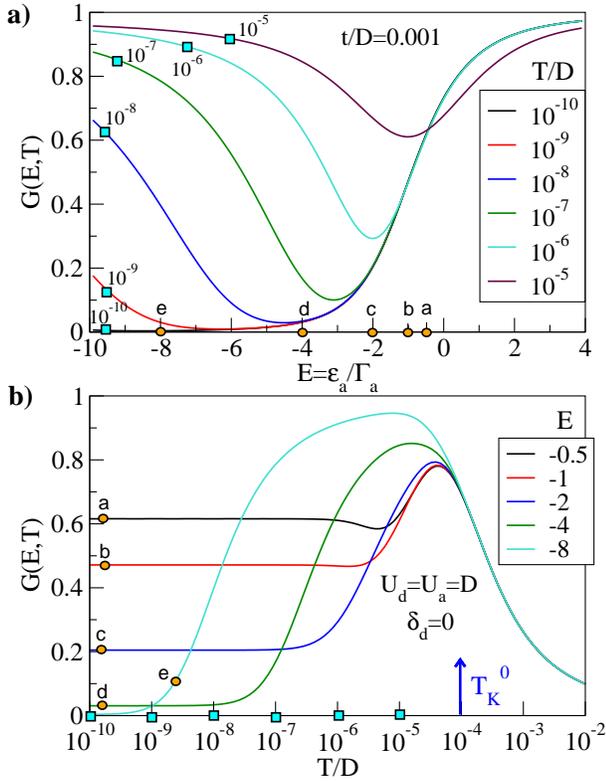}
\caption{(Color online) a) Conductance curves for the fully interacting case
with $U_d/D=U_a/D=1$, $\delta_d=0$, for a range of different temperatures.
b) Temperature dependence of the conductance for a range of energies of the
side-coupled dot.}
\label{fig7}
\end{figure}

Since the experiment is performed at finite temperatures, the results for
the zero-temperature conductance can be very misleading in problems with
very low energy scales \cite{trikotnik}. This commonly occurs in the
side-coupling geometry and in clusters of quantum dot with weak inter-dot
tunnel coupling \cite{cornaglia2005tsk, sidecoupled, flnfl3, trikotnik}. In
Fig.~\ref{fig7}a) we thus plot the conductance for a range of finite
temperatures, which reveals completely different transport properties when
compared with the zero-temperature limit.

When both quantum dots are interacting and the parameters are set in
such a way that local moments form on both dots, there is a
possibility that the Kondo screening will proceed in two steps
\cite{vojta2002, cornaglia2005, sidecoupled, sidecoupled2, flnfl3,
trikotnik, anda2008}: at higher Kondo temperature $T_K^0$, the moment
on the first dot will be screened, while another stage of Kondo
screening of the moment on the side-coupled second dot occurs at
another much lower energy scale $T_K^2$ \cite{cornaglia2005,
sidecoupled}:
\begin{equation}
T_K^2 = c_1 T_K^0 \exp\left(-c_2 T_K^0/J_\mathrm{eff}\right),
\end{equation}
where $c_1$ and $c_2$ are some constants of the order 1. Such {\sl
two-stage Kondo screening} occurs if the effective Kondo exchange
coupling $J_\mathrm{eff}$ of the spin on the second dot with the
quasiparticles resulting from the first stage is lower than $T_K^0$.
In this case, the second Kondo scale may be pushed to exponentially
low temperature $T_K^2$. The exchange coupling
\begin{equation}
J_\mathrm{eff} = \frac{2t^2}{\epsilon_a-\epsilon_d+U_a}
+ \frac{2t^2}{\epsilon_d-\epsilon_a+U_d}
\end{equation}
depends on the value of $\epsilon_a$ and decreases as $\epsilon_a$ is
reduced, attaining its minimum for $\epsilon_a=\delta_d-U_a/2$. In turn,
this implies that $T_K^2$ becomes very low for small $\epsilon_a$, thus at
the experimentally relevant temperatures the second stage of Kondo screening
does not occur, the conductance suppression does not arise and the
conductance curve takes a form similar to that of the Fano resonance. In
experiments, the temperature is roughly one order of the magnitude lower
than the Kondo temperature; for chosen $t/D=0.001$, we indeed find that
for $T/D=10^{-5} \approx 0.1 T_K^0$, the conductance form takes the form of
a nearly symmetric anti-resonance, in agreement with the experiment. The
minimum of the anti-resonance is located at lower value of
$E=\epsilon_a/\Gamma_a$ as compared to the non-interacting case (where it is
found exactly at $E=0$), since the physical mechanism for its occurrence is
completely different: in the non-interacting case we observe a {\it
complete} suppression of the conductance due to quantum interference between
different conductance pathways, while in the interacting case the {\it
incomplete} suppression results from the separation of the energy scales
$T_K^0$ and $T_K^2$ with the experimental temperature set on an intermediate
scale so that $T_K^2 < T < T_K^0$. The shift from $E=0$ is not observable
experimentally, since the exact value of the parameter $\epsilon_a$ is not
known: it has to be inferred indirectly from known gate voltages on the
electrodes and the capacitances.

\begin{figure}[htbp]
\includegraphics[width=8cm,clip]{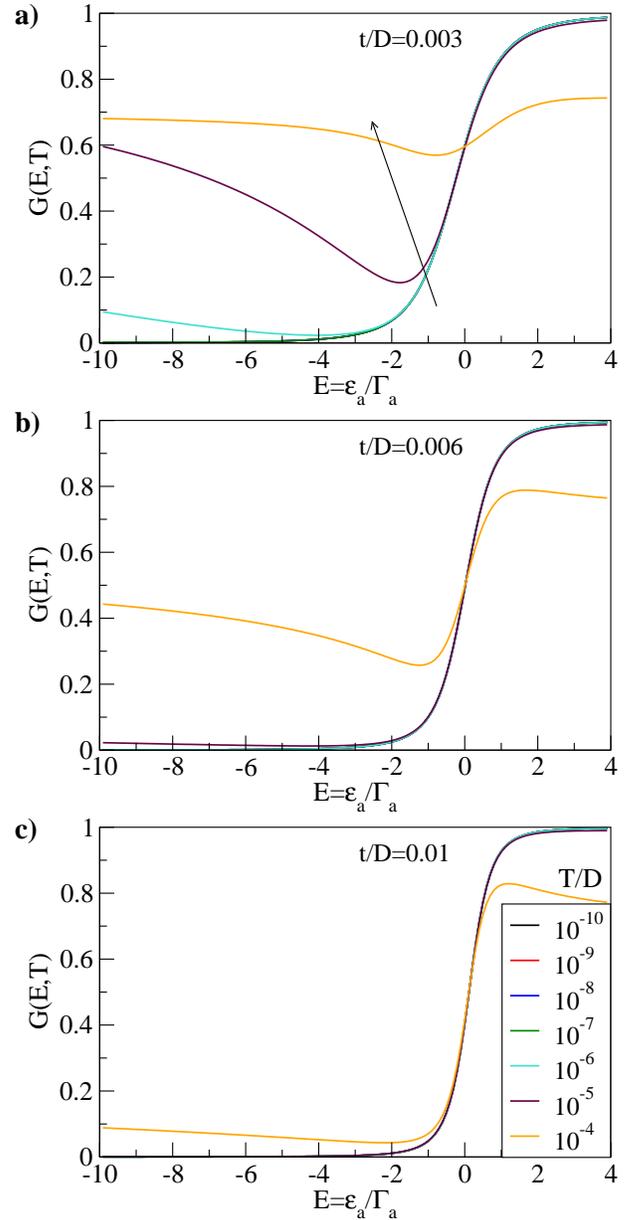}
\caption{(Color online) Conductance curves for the fully interacting case
with $U_d/D=U_a/D=1$, $\delta_d=0$, for increasing inter-dot tunneling
$t$. The arrow indicates the direction of the increasing temperature $T$.}
\label{fig8}
\end{figure}

If the inter-dot tunneling $t$ is increased so that the exchange
coupling becomes comparable or exceeds the Kondo temperature, the
conductance curves no longer exhibit a Fano-like antiresonance, not
even at elevated temperatures, see Fig.~\ref{fig8}. Instead, as $E$ is
reduced so that the side-coupled dot becomes occupied by one electron,
the two electrons (one from each quantum dot) bind to a local singlet
on the temperature scale of $J \sim 4t^2/U$ and there are no other
lower temperature scales.

The interpretation of the anti-resonance in terms of a thermally
suppressed second-stage Kondo screening is supported by the
experimental results for the temperature-dependence of the resonant
conductance (Fig. 2c in Ref.~\onlinecite{sasaki2009}) which indicate a
logarithmic behavior. This temperature variation corresponds to the
high-temperature range in Fig.~\ref{fig7}b). The experimental results
do not indicate any saturation of $G$ for the lowest experimentally
accessible temperatures, however this might simply indicate that the
temperature is not sufficiently below $T_K$, or that $t$ is indeed
very small and thus $T_K^2$ is exponentially low. Further confirmation
for the interpretation in terms of the two-stage Kondo effect could be
obtained by also considering the temperature-dependence of other
parameters which determine the anti-resonance line-shape; the
variations of amplitude, width, and asymmetry parameter should all
exhibit logarithmic temperature dependence.

In addition to the thermal effects, the second stage of Kondo screening may
also be suppressed by a minute magnetic field such that $T_K^2 \ll B \ll
T_K^0$. As it is likely that there are always some small stray magnetic
fields in the experimental device, this provides another mechanism which
tends to produce Fano-resonance-like spectral features in the present
configuration.

\section{Conclusion}

The transport properties of a system of two quantum dots in the side-coupled
geometry have been analyzed, focusing on the parameter range where the
occupancy of the side-coupled dot changes as its energy crosses the Fermi
level. The resulting resonance line-shapes in the conductance curves have
been comprehensively studied in various parameter regimes for both
non-interacting and interacting quantum dots. We find an alternative
interpretation of the experimental results from Ref.~\onlinecite{sasaki2009}
which fulfils the two required conditions: 1) the occupancy in the
side-coupled dot changes by one, 2) the resonance line-shape features a
weakly asymmetric Fano-like anti-resonance. The proposed model takes into
account the electron-electron interaction on the side-coupled dot, which is
certainly present in the experimental device and which has been shown in
this work to be relevant for the transport properties of the system.

\begin{acknowledgments}
The author acknowledges the support of the Slovenian Research Agency (ARRS)
under Grant No. Z1-2058.
\end{acknowledgments}

\bibliography{paper}

\end{document}